\documentclass{aa}
\usepackage{times,amsmath,amssymb,txfonts,graphicx,aas_macros,natbib}
\usepackage{xspace,color,multirow}

\def\pasa{{\rm PASA}}

\newcommand{\Eq}[1]{Eqn.~(\ref{#1})}

\newcommand{\Eqss}[2]{Eqns.~(\ref{#1}) -- (\ref{#2})}

\newcommand{\Sec}[1]{Sect.~\ref{#1}}

\newcommand{\Fig}[1]{Fig.~\ref{#1}}

\newcommand{\Tab}[1]{Table~\ref{#1}}

\usepackage{ulem}

\renewcommand\emph[1]{\textit{#1}}

\usepackage{relsize}
\newcommand\HI{H{\smaller I}\xspace}
\newcommand\HII{H{\smaller II}\xspace}

\newcommand\Myr{\,\rm Myr}
\newcommand\Gyr{\,\rm Gyr}

\newcommand\kms{\,\rm km\,s^{-1}}
\newcommand\pc{\,\rm pc}

\newcommand\muG{\,\mu\rm G}
\newcommand\nG{\,\rm nG}
\newcommand\kpc{\,\rm kpc}

\newcommand\Msun{\,\rm M_\odot}

\newcommand\yes{$\bullet$}
\newcommand\no{$\circ$}

\newcommand\tms{\!\times\!}
\newcommand\cdt{\!\cdot\!}

\newcommand\mrho{\bar{\varrho}}
\newcommand\V{\bar{\mathbf v}}
\newcommand\B{\bar{\mathbf B}}

\newcommand\mB{\mn{\mathbf{B}}}
\newcommand\EMF{\bar{\mbox{\boldmath{${\cal E}$}}}}

\newcommand{\mn}[1]{\overline{#1}}

\newcommand\ra{$\rightarrow$}

\newcommand\Pm{\mathrm{Pm}}

\newcommand\visfl{\boldsymbol{\tau}}
\newcommand\etat{\eta_{\rm t}}
\newcommand\vist{\nu_{\rm t}}

\newcommand{\simgt}%
           {\,\hbox{\lower0.35ex\hbox{$\sim$}\llap{\raise0.35ex\hbox{$>$}}}\,}
\newcommand{\simlt}%
           {\,\hbox{\lower0.35ex\hbox{$\sim$}\llap{\raise0.35ex\hbox{$<$}}}\,}

\newcommand\NIII{\textsc{nirvana-iii}\xspace}

\begin{document}

\title{Toward a hybrid dynamo model for the Milky Way}
\titlerunning{Hybrid galactic dynamo models}

\author{Oliver~Gressel\inst{1} \and Detlef~Elstner\inst{2} 
  \and Udo Ziegler\inst{2}}

\offprints{O. Gressel, \email{gressel@kth.se}}

\institute{NORDITA, KTH Royal Institute of Technology and 
  Stockholm University, Roslagstullsbacken 23, 106 91 Stockholm,
  Sweden \and Leibniz-Institut f{\"u}r Astrophysik Potsdam (AIP), 
  An der Sternwarte 16, 14482 Potsdam, Germany}

\date{Received <date> / Accepted <date>\vskip-4ex}


\abstract
{Based on the rapidly increasing all-sky data of Faraday rotation
measures and polarised synchrotron radiation, the Milky Way's magnetic
field is now modelled with an unprecedented level of detail and complexity.}
{We aim to complement this heuristic approach with a physically
motivated, quantitative Galactic dynamo model -- a model that moreover
allows for the evolution of the system as a whole, instead of just
solving the induction equation for a fixed static disc.}
{Building on the framework of mean-field magnetohydrodynamics and
extending it to the realm of a \emph{hybrid} evolution, we perform
three-dimensional global simulations of the Galactic disc. To
eliminate free parameters, closure coefficients embodying the
mean-field dynamo are calibrated against resolved box simulations of
supernova-driven interstellar turbulence.}
{The emerging dynamo solutions comprise a mixture of the dominant
axisymmetric S0 mode, with even parity, and a sub-dominant A0 mode,
with odd parity. Notably, such a superposition of modes creates a
strong localised vertical field on one side of the Galactic disc. We
moreover find significant radial pitch angles, which decay with radius
-- explained by flaring of the disc. In accordance with previous work,
magnetic instabilities appear to be restricted to the less-stirred
outer Galactic disc. Their main effect is to create strong fields at
large radii such that the radial scale length of the magnetic field
increases from $4\kpc$ (for the case of a mean-field dynamo alone) to
about $10\kpc$ in the hybrid models -- the latter being in much better
agreement with observations.}
{There remain aspects (e.g., spiral arms, X-shaped halo fields,
fluctuating fields) that are not captured by the current model and
that will require further development towards a fully dynamical
evolution. Nevertheless, the work presented demonstrates that a hybrid
modelling of the Galactic dynamo is feasible and can serve as a
foundation for future efforts.}

\maketitle


\section{Introduction}
\label{sec:intro}

The Galactic magnetic field (GMF) can now be modelled with an ever
increasing level of detail \citep[see e.g.][]{2007ApJ...663..258B,%
  2010MNRAS.401.1013J,2011MNRAS.416.1152J,2011A&A...526A.145F,%
  2011ApJ...728...97V,2012ApJ...757...14J,2012ApJ...761L..11J,%
  2012ApJ...755...21M}. This becomes possible with the availability of
all-sky data of Faraday rotation measures \citep{2012A&A...542A..93O}
and polarised synchrotron emission obtained by space missions like
WMAP or Planck \citep[see][]{2012A&A...540A.122F}. The typical
approach utilises $\chi^2$ minimisation for fitting a large number of
free parameters.

Theoretical models for the GMF are largely heuristic and guided by
existing knowledge, e.g. derived from external galaxies. Modelling
assumptions such as the winding angle of the spiral arms and its
variation with radius (and near reversals) remain under debate
\citep[see e.g. discussion in][]{2010ASPC..438..216B}. From a
theoretician's point of view, these models still provide a convenient
link to observations and can serve as a benchmark for dynamo models in
the framework of mean-field MHD
\citep{1996ARA&A..34..155B,2012SSRv..tmp...57B}. Despite the success
of the heuristic description, a thorough understanding of the
underlying field amplification mechanism appears desirable.

Different flavors of mean field models have been studied for typical
galaxies, yet under many simplifying assumptions. As for models
specifically designed for the GMF, the most prominent difference of
the Milky Way magnetic field in comparison with external galaxies
shows up in the observationally now well supported \citep[see
  e.g. discussion in][]{2011PASA...28..171K} reversal of the mean
magnetic field in the radial direction -- which has never yet been
observed in other galaxies.\footnote{It should however be mentioned
  that we can only expect to detect field reversals unambiguously in a
  limited number of nearby galaxies.}  Mean field models allow such
field reversals in principle, depending on the seeding of the dynamo
process \citep{2012A&AT...27..319M}. Alternative explanations include
oscillating solutions due to a vertical dependence of differential
rotation \citep{2000A&A...358..125F} or, as proposed in this paper, a
vertical undulation of the Galactic midplane combined with an
antisymmetric vertical parity of the disc field, leading to apparent
reversals. Yet another possibility, investigated here for the first
time, is the mode interface between the dynamo-dominated inner region
and the instability-dominated outer region of the Galaxy.

A further peculiarity of the GMF is the rather small pitch angle (of
$\simlt 10\degr$) compared to many other galaxies with similar strong
differential rotation and pitch angles up to $\sim45\degr$ like, e.g.,
observed in M94 and M33
\citep{2008ApJ...677L..17C,2008A&A...490.1005T}.  This dominance of
the azimuthal field appears to be in better agreement with mean field
models than with the large pitch angles observed in external galaxies.
The magnetic field of mean-field dynamos is usually a stationary
axisymmetric quadrupole. But for some cases like weak differential
rotation or including the halo in the dynamo process, dipolar and/or
oscillatory solutions may occur
\citep{1992A&AS...94..587E,1992A&A...259..453B,2010A&A...512A..61M}.
The flaring of the Galactic disc was usually ignored in the modelling
but has been considered in some recent publications
\citep{2013A&A...556A.147M,2013MNRAS.428.3569C,2013MNRAS.433.3274C}. This
is despite the fact that observations seem to favour a non-flaring
disc, at least in \HII -- see \citet{1998ApJ...497..238L}, and
discussion in \citet{2013A&A...556A.147M}. Our own model is guided by
more recent observations \citep{2008A&A...487..951K} of the \HI
distribution, which is indeed found to be flared.

A notable exception to dynamo models with constant scale height is the
work by \cite*{1993MNRAS.264..285P}, who assume a flared gas
distribution and apply mean field models in the thin disc
approximation to the Milky Way, relying on an observed rotation curve,
and estimates for disc height, turbulent velocity, correlation length
and gas density. They derive radial profiles of the final magnetic
field for regular and chaotic seed fields with $\alpha$~quenching due
to magnetic helicity. The maximum field strength of the regular field
appeared at about $6\kpc$. For strong enough seed field (0.001 to
$0.1\muG$) they found reversals.

While such models have provided us with a wealth of qualitative
understanding of the expected mode structure appearing in
$\alpha\Omega$~type disc dynamos, they lack a rigorous foundation for
the actual amplitude and spatial distribution of the imposed
mean-field effects. Recently, a quantitative measurement of the
transport coefficients has become possible by employing the so-called
test-field (TF) method \citep{2005AN....326..245S} to realistic local
simulations of interstellar turbulence
\citep{2008AN....329..619G}. This also includes the determination of
quenching functions \citep*{2013MNRAS.429..967G}, required to evolve
the mean-field models into the saturated regime. Based on this
previous work, we here present a quantitative mean-field dynamo
model. By doing so, we aim to provide a comprehensive description of
the GMF which is backed-up by observable properties of the Galaxy.

A further improvement over existing work is the \emph{combined}
evolution of the induction and momentum equations. While there exist
global three-dimensional MHD simulations of galactic gaseous discs
\citep[e.g.][]{2004A&A...423L..29D,2009ApJ...706L.155H,%
  2011ApJ...733L..18K,0004-637X-764-1-81}, these simulations typically
do not account for the effects caused by the vigorous SNe turbulence
on scales unresolved on the global mesh. On the other hand, in
classical mean-field models, the velocity field is kept fixed at its
initially prescribed state. In contrast to this, the inclusion of the
momentum equation allows the underlying disc model to evolve in
time. At the same time, solving the full MHD equations (subject to
parametrised \emph{enhanced} dissipation) permits for the emergence of
magnetic instabilities like the magnetorotational instability (MRI)
and buoyancy instabilities.

Clearly, the model presented still lacks important aspects of the
evolution of the galactic disc: it ignores a self-consistent
prescription of e.g. self-regulatory star formation (SF), formation of
spiral arms via self-gravity, emergence of a Galactic wind driven by a
cosmic ray (CR) component, or the multi-phase nature of the
interstellar medium (ISM). Nevertheless, the presented framework can
be regarded as a first step towards less static dynamo models.


\section{Methods}
\label{sec:methods}

As outlined above, we aim to derive a quantitative model for the
Galactic dynamo which is based, as directly as possible, on observable
quantities. Rather than solving the induction equation with a given
static velocity field, $\V$, we here intend to perform simulations of
the full mean-field MHD equations
\begin{eqnarray}
  \partial_t\mrho +\nabla\cdt(\mrho\,\V) & = & 0
  \,,\label{eq:continuity}\\[2pt]
  \partial_t(\mrho\,\V) +\nabla\cdot
          \left[\;\mrho\,\V\,\V+\bar{p}^{\star}-\B\B
          \;\right] & = &
          - \mrho\,\nabla\Phi + \nabla\cdt\visfl
          \,,\label{eq:mom}\\[2pt]
  \partial_t \B - \nabla \times \left[\;\V\tms\B + \EMF
                - \etat \nabla\tms\B \;\right] & = & 0 \,,\label{eq:ind}
\end{eqnarray}
where the viscous stress tensor is given by $\visfl = \mrho\,\vist \,
[\nabla\V + (\nabla\V)^{\top} -2/3\,(\nabla\cdt\V)\,I ]$ with
kinematic viscosity $\vist$, and $\bar{p}^{\star}$ denotes the total
pressure $\bar{p}+\bar{B}^2/2$. As made explicit by writing vertical
bars over the constituent variables, \Eqss{eq:continuity}{eq:ind} are
mean-field equations governing the evolution of \emph{large-scale}
quantities. We want to emphasise that the given set of equations still
neglects a number of additional ``micro-physics'' which should be
considered in the future (see \Sec{sec:negl} for a discussion). As
justified by the immense Reynolds numbers within the ISM, we
furthermore ignore contributions to the dissipation coefficients
stemming from molecular effects. In this sense, $\etat$, and $\vist$
represent the turbulent diffusivity and kinematic viscosity,
respectively. Induction effects stemming from unresolved scales, are
included in \Eq{eq:ind} in the form of a mean electromotive force,
$\EMF$, which we specify in \Sec{sec:emf}. We have modified the
publicly available \NIII code
\citep{2004JCoPh.196..393Z,2011JCoPh.230.1035Z} to include this
additional EMF and have verified our implementation against the
benchmark described in \citet{2008A&A...483..949J}.

For the simulations presented in this paper, we chose a domain size of
$r\in[1.5,21.5]\kpc$, $\theta\in[0.415,0.585]\,\pi$, and
$\phi\in[0,2]\,\pi$. Note that, owing to the use of polar coordinates,
the central region of the disc is excluded for reasons of
computational expedience. We typically employ a resolution of
$256\times 48\times 64$ grid points in the radial, latitudinal, and
azimuthal directions, respectively. For the static 2D simulations, we
have further checked convergence when increasing the resolution to
$512\times 96$ grid points. For the fully dynamic 3D simulations we
ran simulations up to $384\times 72 \times 96$. The hydrodynamic
boundary conditions (BCs) are of the standard `outflow' type; for the
lower $\theta$~boundary this, e.g., implies setting $\partial_\theta
v_\theta=0$ if $v_\theta<0$, and enforcing $v_\theta=0$ otherwise. We
furthermore solve for hydro\-static/dynamic balance in the vertical
and radial directions, respectively, and impose the equilibrium
rotation profile at the inner radial boundary. For the magnetic field
we employ both `pseudo-vacuum' (i.e., $B_\parallel=E_\perp=0$) and
`perfect-conductor' ($B_\perp=E_\parallel=0$) boundary conditions. We
moreover impose the initial net-vertical field on the boundaries if
present.

In the following, we begin our formulation by specifying a complete
disc model. We will then prescribe turbulent closure coefficients
subsuming effects due to the turbulence driven by supernovae (SNe).


\subsection{Equilibrium disc model}

The initial density profile and rotation curve are constructed to be
in a stationary hydrodynamic equilibrium with a given gravitational
potential, consisting of a standard
\citeauthor*{1997ApJ...490..493N}~(NFW) dark-matter halo, a component
due to a stellar disc \citep{1975PASJ...27..533M}, and a central
bulge. We chose these simplified prescriptions because they are
commonly used for fitting observational data, allowing reasonable
constraint of the shape parameters. For simplicity, we ignore the
contributions due to the central bar, and the self-gravity of the
gaseous disc.

\subsubsection{External gravity}

The gravitational potential due to the Galactic dark-matter halo can
be approximated by \citep*{1997ApJ...490..493N}:
\begin{equation}
  \Phi_{\rm DM} = -\frac{G\,M_{\rm H}}{R_{\rm H}}\,g(c)
  \frac{\ln\,(1-cx)}{x}\,,\quad x\equiv r/R_{\rm H}
\end{equation}
with $r$ the spherical radius, $g(c)$ the NFW shape function, and
where we have chosen a concentration parameter of $c=13$, and $R_{\rm
  H}=213\kpc$, and $M_{\rm H}= 10^{12}\Msun$ are the assumed virial
radius, and virial mass of the Milky Way dark-matter halo
\citep[cf.][]{2008ApJ...684.1143X}, respectively. For the stellar
disc, we assume a parametrisation according to
\citet{1975PASJ...27..533M}, with
\begin{equation}
  \Phi_{\rm \star} = \frac{G\,M_{\star}}%
                          {\sqrt{R^2+\left[a+\sqrt{z^2+b^2}\right]^2}}\,,
  \label{eq:MiyNag}
\end{equation}
where, $R\equiv r\,\sin(\vartheta)$ is now the cylindrical radius, and
$z\equiv r\,\cos(\vartheta)$ is the vertical coordinate. We adopt a
total mass of $M_{\rm disc}=7\times10^{10}\Msun$, and shape parameters
$a=3.5\kpc$, and $b=0.18\kpc$ in accordance with the potential used in
our local box simulations situated at $R=8.5\kpc$
\citep[cf.][]{2008AN....329..619G}. For a simplified representation of
a Galactic bulge \citep{1996MNRAS.281.1027F} with an assumed mass
$M_{\rm bulge}=1.6\times10^{10}\Msun$, we again use the expression
(\ref{eq:MiyNag}), but now with $a=0\kpc$, and $b=0.42\kpc$, resulting
in a spherically symmetric potential. Ignoring the self-gravity of the
gas, the effective gravitational potential we use is given by
$\Phi(R,z) \equiv \Phi_{\rm DM} + \Phi_{\rm disc} +\Phi_{\rm bulge}$.

\subsubsection{Disc model and rotation curve}

\begin{figure}
  \includegraphics[width=\columnwidth]{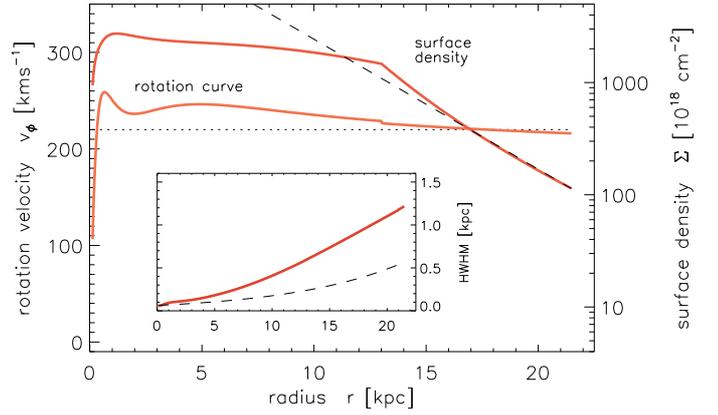}
  \caption{Midplane rotation curve (dotted line $=220\kms$) and disc
    surface density. The inset shows the flaring scale-height of the
    \HI gas disc. Dashed lines indicate observational approximations
    suggested by \citet{2008A&A...487..951K}.}
  \label{fig:disc_prof}
\end{figure}

Our hydrodynamic model is based on the flaring \HI disc of the Milky
Way, which has recently been constrained observationally by
\citet{2008A&A...487..951K}. Improving their simple exponential fit
for the radial profile of the surface density, we propose a split
profile with a radial break, yielding\footnote{For reasons of compact
  notation we here use various coordinate systems. All simulations are
  performed in a spherical-polar mesh ($r$,$\vartheta$,$\phi$).}
\begin{equation}
  \mrho(R,z) = \mrho(R_{\rm br},0)\;{\rm e}^{-\Phi_z/c_{\rm s}^2}\;\times\;
  \begin{cases} 
    \;\exp ( -\frac{R-R_{\rm br}}{R_{\rm exp}} ) 
    & \text{for $R<R_{\rm br}$} \\[4pt]
    \;\left(\frac{R}{R_{\rm br}}\right)^p
    & \text{for $R\ge R_{\rm br}$} 
  \end{cases}
  \label{eq:rho_mid}
\end{equation}
for the initial density distribution in hydrostatic equilibrium. Here
the vertical disc structure is given by $\exp(-\Phi_z/c_{\rm s}^2)$,
i.e., by setting the thermal pressure in balance with the function
\begin{equation}
  \Phi_z(R,z) \equiv \Phi(R,z) - \Phi(R,z=0)\,,
\end{equation}
expressing the vertical potential difference
\citep[cf.][]{2010MNRAS.407..705W}. Furthermore, the reference density
$\mrho(R_{\rm br},0)$ is specified at the break radius $R_{\rm
  br}=13\kpc$, and we use $R_{\rm exp}=4\kpc$, which is somewhat less
steep than the $3.75\kpc$ proposed by
\citeauthor{2008A&A...487..951K}. For our fiducial model, we chose an
exponent $p=-6.5$, and we note that the power-law closely reproduces
the upward curvature seen in figs.~3-6 of \citet{2008A&A...487..951K},
and which cannot be matched by a single exponential. The disc surface
density of our model is shown in \Fig{fig:disc_prof}, where we also
plot the flaring scale height, $h(R)$, of the disc. The flaring of the
disc is controlled by prescribing a radial temperature profile such
that
\begin{equation}
  c_{\rm s}^2(R,z) = c_{{\rm s}0}^2\, \left(\frac{R}{R_0}\right)^q
\end{equation}
is a power-law function of $R$. To be more specific, a value $q=-1$
would, e.g., produce a disc with constant opening angle, whereas
$q\simeq 0$ would lead to a globally isothermal, flaring disc. Limited
by numerical feasibility, we here chose $q=-0.5$ and $h/R=0.15$ (at
$R_0=10\kpc$), which leads to a more inflated disc (see inset
\Fig{fig:disc_prof}) but still provides a reasonable fit to the
observed $h(R)$ -- also cf. fig.~7 in \citet{2008A&A...487..951K}. A
discussion of the effect of a radially non-uniform scale height on the
dynamo can, e.g., be found in section VII.6 of
\citet{1988ASSL..133.....R}.

We remark that the functional form of (\ref{eq:rho_mid}) is not
differentiable at $R_{\rm br}$, which implies a slight jump in the
azimuthal velocity, which we derive as
\begin{equation}
  \bar{v}_{\phi}^2(R,z) =  \left(
  \left.\frac{\partial \ln \mrho}{\partial \ln R}\right|_{z=0} 
  \!+ q\, \right)\, c_{\rm s}^2
  + R\, \left.\frac{\partial \Phi}{\partial R}\right|_{z=0}
  + q\,\Phi_z(R,z)\,.
  \label{eq:v_phi}
\end{equation}
We note that, for $q\ne 0$, the last term will lead to a vertical
variation of the rotation profile. Because the prescribed
gravitational potential has been designed as a fit to the Galactic
rotation curve, it suffices to say that (\ref{eq:v_phi}) reproduces
the classical Brandt-type curve with a plateau at
$\bar{v}_{\phi}\simeq 220\kms$ as also shown in \Fig{fig:disc_prof},
where we plot the rotation curve in the disc midplane. This completes
the description of our hydrodynamic disc model. We have checked that
the inviscid model can be evolved stably for times exceeding the age
of the universe. When including turbulent viscosity, the disc evolves
viscously on a secular time scale.


\subsection{Turbulent closure parameters}
\label{sec:meanfield}

Utilising a mean-field approach, we aim to simulate the common
evolution of the regular magnetic field \emph{and} the large-scale
flow of the Galactic gas disc. Conceptually, this implies that we
solve the full MHD equations subject to turbulent transport
coefficients, i.e., we prescribe a turbulent diffusivity in the
induction equation along with an according turbulent viscosity in the
momentum equation. These turbulent dissipation coefficients, as well
as the prescribed $\alpha$~effect (see \Sec{sec:emf}), are derived
from a series of \emph{resolved} direct numerical simulations
(DNS). The basic setup, which assumes a local box geometry and uses
realistic driving of interstellar turbulence via injection of
localised supernova explosions, is described in detail in
\citet{2009PhDT........99G}. Scaling relations for mean-field effects
have been derived from a set of models with box size
$0.8\kpc\times0.8\kpc\times\pm 2.1\kpc$, and more accurate vertical
shapes have been obtained using a larger $1.6\kpc\times 1.6\kpc\times
\pm 6.4\kpc$ domain \citep{2011IAUS..274..348G}. The requirement to
resolve individual explosions demands grid resolutions below $10\pc$,
which is met in the box simulations but will remain unaffordable in
global simulations for some time.

Since in the DNS we only measure $\etat$, but not $\vist$, we have to
assume a \emph{turbulent} magnetic Prandtl number, $\Pm_{\rm
  t}=\vist/\etat$ of unity. This assumption is backed by recent
numerical investigations that found very little deviation from this
value \citep{2009ApJ...697.1901G,2009A&A...507...19F}. Note that
unlike for a classical ``$\alpha$''~viscosity, we prescribe
$\vist(R,z)$ directly as a function of space, i.e., independent of the
gas pressure. This is because (for consistency with the mean induction
equation) we assume the turbulent velocity field to be given \emph{a
  priori} as a consequence of the underlying SNe distribution. To
avoid restrictive time-step constraints arising from high values of
the diffusion coefficients, we have implemented the
super-time-stepping scheme, introduced by \citet*{CNM:CNM950}, for the
viscous and diffusive updates.

\begin{figure*}
  \includegraphics[width=1.9\columnwidth]{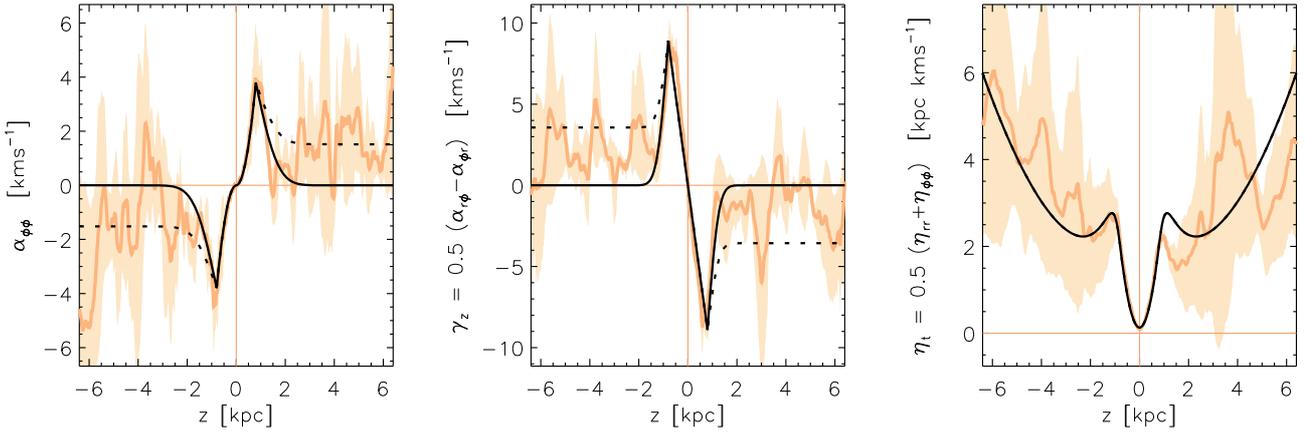}
  \caption{Vertical profiles of the TF coefficients
    $\alpha_{\phi\phi}$ (left), $\gamma_z$ (centre), and $\eta_{\rm
      t}$ (right panel) measured from DNS (light colours); shaded
    areas indicate $1\sigma$ fluctuations. Black curves show model
    profiles $\hat{\alpha}(\zeta)$, $\hat{\gamma}(\zeta)$, and
    $\hat{\eta}(\zeta)$ used for the simulation \emph{with} (dashed
    line) and \emph{without} (solid line) a halo dynamo.}
  \label{fig:ae_prof}
\end{figure*}

\subsubsection{Neglected effects}
\label{sec:negl}

Turbulent viscosity is by no means the only mean-field effect in the
momentum equation but, in fact, a crude oversimplification. For
reasons of tractability, we however have to ignore more involved
contributions like, e.g. the turbulent kinetic pressure or the
turbulent contribution to the Lorentz force in our current
considerations. A rudimentary attempt to allow the former effect to
self-consistently launch a Galactic wind (driven by the gradient in
the turbulence intensity) led to unacceptable mass-loss rates. This is
because, in our mean-field model, the single-phase density field
cannot properly capture the multi-phase nature of the ISM: Ultimately,
what we described as a ``wind'' is really a ``fountain flow'', i.e.,
tenuous gas being blown out of the Galaxy, while high-density clumps
raining down compensate the overall mass balance. Ultimately, it will
be of great interest to study related effects. We speculate that the
suppression of the turbulent magnetic pressure by mean-fields
\citep{1989SvAL...15..274K,2012ApJ...749..179B} may also be of
importance in a Galactic context, potentially leading to an increased
heterogeneity of the observed field \citep[but also
  see][]{2009IAUS..259...87F}.

To be able to use periodic boundary conditions, an important
simplification of our local box simulations was to ignore any
large-scale radial structure in, e.g., the gas density or supernova
rate. Even though we were able to vary quantities like the mid-plane
density, rotation rate, or shearing rate for each of the individual
runs (i.e. to adjust to the situation at different locations within
the Galaxy), we could not obtain contributions in the dynamo tensor
due to radial \emph{gradients}.


\subsubsection{The dynamo tensor}
\label{sec:emf}

We here focus on a well-studied effect appearing in the mean induction
equation, i.e., the turbulent electromotive force
$\EMF=\overline{\mathbf{v'\tms B'}}$, where primes denote fluctuating
quantities, and which we implemented by means of the classic tensor
prescription
\begin{equation}
  \EMF_i = \alpha_{ij}\,\bar{B}_j\,
  \qquad i,j\,\in\left[r,\vartheta,\phi\right]
\end{equation}
i.e., with an $\alpha$~tensor locally relating the mean EMF to the
mean magnetic field. We argue that the good quantitative agreement
between DNS and one-dimensional mean-field simulations
\citep{2009PhDT........99G} warrants the neglect of non-local
\citep{2008A&A...482..739B} as well as non-instantaneous
\citep{2009ApJ...706..712H} contributions to the closure. As laid out
below, $\alpha_{ij}$ is parametrised according to coefficients
measured within a comprehensive set of DNS of SN-driven ISM turbulence
by means of the TF method \citep{2005AN....326..245S,2007GApFD.101...81S}.

Because of our flaring disc model, and for reasons of simplicity, we
identify the $x$, $y$, and $z$ direction in these local Cartesian box
simulations \citep{2008A&A...486L..35G,2009IAUS..259...81G} with
spherical polar coordinates $r$, $\phi$ (azimuth), and $\vartheta$
(co-latitude), respectively. To reflect the geometry of the disc,
shape functions are defined with respect to a ``flaring'' coordinate
$\zeta \equiv z/h(r)$, as to follow the local scale-height of the \HI
gas disc. At any spherical radius, $r$, the latitudinal variation is
approximated by the profiles shown in \Fig{fig:ae_prof}. Note that (in
accordance with our box model) the aspect ratio, $h(R)/R$, for the
vertical profiles of the dynamo tensor is about a factor of two larger
than the one for the \HI disc shown in the inset of \Fig{fig:ae_prof},
resulting in a scale height of $\sim 1\kpc$ at $R\simeq 10\kpc$.


\begin{table}\begin{center}
\caption{Scaling exponents for the dynamo $\alpha$~tensor, the
  turbulent diffusivity and viscosity, and vertical fountain flow,
  $\mn{v}_z(\zeta)$. Amplitudes refer to shape functions
  $\hat{\alpha}(\zeta)$, $\hat{\gamma}(\zeta)$, and
  $\hat{\eta}(\zeta)$, which we matched to the TF profiles shown in
  \Fig{fig:ae_prof} \citep*[also cf.][]{2011IAUS..274..348G}.
\label{tab:coeff}}
\begin{tabular}{lrlccccc}\hline
 & \multicolumn{3}{c}{amplitude}
 $\sigma/\sigma_0$ & $\Omega/\Omega_0$ & $\mrho/\mrho_0$ \\[4pt]
 \hline
 $\alpha_{\phi\phi}$ & $  2$ & $\kms$ & 0.4 & 0.5& -0.1 \\
 $\alpha_{\phi r}\,,-\alpha_{r\phi}$ & $-12.5$ &$\kms$ & 0.45 & -0.2 & 0.3 \\
 $\mn{v}_z$ & 8 & $\kms$ & 0.4 & -- & -- \\
 $\etat\,,\vist$ & $2$ & $\kpc\kms$ & 0.4 & 0.25$\,^a$\hspace{-2ex} & 0.4 \\
 \hline
\end{tabular}
\end{center} \footnotesize $^a$) revised exponent from new analysis
\end{table}


The tensor components $\alpha_{rr} \simeq \alpha_{\phi\phi} \simeq
5\,\alpha_{\theta\theta}$, and $\alpha_{r\phi}\simeq-\alpha_{\phi r}$,
which we directly measured in the DNS, are parametrised according to
their inferred dependence on the supernova rate $\sigma/\sigma_0$,
rotation rate $\Omega/\Omega_0$, and disc midplane density
$\mrho/\mrho_0$. To further reduce the number of free parameters, we
identify $\sigma/\sigma_0$ with the star formation rate, and link it
to the local surface density by means of a Kennicutt-Schmidt law
\begin{equation}
  \sigma/\sigma_0 = (\mrho/\mrho_0)^{1.4}
\end{equation}
\citep{1998ApJ...498..541K}. Scaling exponents are listed for
reference in \Tab{tab:coeff}. Note that, for reasons of completeness,
we include the radial and vertical diagonal elements of the tensor,
even though we find these terms to be negligible compared to the
$\alpha\Omega$ mechanism driven via $\alpha_{\phi\phi}$.\footnote{Also
  note that \citet{2008A&A...486L..35G} found the $\alpha^2$~dynamo to
  be only marginally excited for the coefficients measured in DNS.}
Another important result from that paper was the existence of a
vertical fountain flow, $\mn{v}_z$, which was discerned to balance the
effect of the turbulent pumping. The vertical profile
$\mn{v}_z(\zeta)$ which we adopt in our mean field prescription
consists of two parts: a contribution linear in $\zeta$, and a
characteristic modulation \citep[cf. figure~2
  in][]{2009IAUS..259...81G}, which roughly coincides with
$\alpha_{r\phi}(\zeta)$. It is important to note that the mean flow
$\mn{v}_z$ -- for reasons discussed in \Sec{sec:negl} -- only
contributes to the induction equation but is \emph{not} included in
the momentum equation.

\subsubsection{Quenching functions}

Avoiding a more complex description relating to the (approximate)
conservation of magnetic helicity \citep[see e.g. discussion
  in][]{2011AN....332...88M}, we here resort to a simple algebraic
expression for the $\alpha$~quenching as justified by our recent
analysis \citep*{2013MNRAS.429..967G}. Unlike in many earlier studies,
we explicitly include a quenching for the $\etat$ (and $\vist$)
coefficient \citep[also see][]{2003A&A...411..321Y}. Adopting an
isotropic quenching for the $\alpha$~tensor, we use
\begin{equation}
  \alpha_{ij}(\bar{B}) = \frac{\alpha_{ij}\,(\bar{B}\!=\!0)}%
                         {1+q_{\alpha}\,\beta^2}\,,
  \quad
  \vist(\bar{B}) =
  \etat(\bar{B}) = \frac{\etat(\bar{B}\!=\!0)}%
                         {1+q_{\eta}\,\beta}
  \label{eq:quen}
\end{equation}
with $\beta \equiv |\bar{B}|/B_{\rm eq}$, and coefficients
$q_\alpha=10$, and $q_\eta=5$ approximating the results from direct
simulations \citep{2013MNRAS.429..967G}. Because of the close
correspondence of $\alpha_{r\phi}$ and the characteristic modulation
seen in the mean flow \citep[cf. figure~3b in][]{2013MNRAS.429..967G},
we chose to apply quenching only to the modulation in $\bar{v}_z$, and
leave the underlying linear profile unquenched.

For consistency, we moreover compute the equipartition field strength
$B_{\rm eq}^2\equiv \mu_0\mrho\,v_{\rm rms}^2$ from a turbulent
velocity profile $v_{\rm rms}(\zeta)$ which has itself been derived
from the (unquenched) $\etat(\zeta)$ profile, assuming the classical
relation $\etat=\frac{1}{3}\tau_{\rm c}\,v_{\rm rms}^2$, and where we
have used a constant $\tau_{\rm c}=3.5\Myr$ in consistence with DNS.

Note that it is essential that the $\alpha$~effect and the turbulent
diffusion are quenched differently. This becomes obvious by evaluating
$C_\alpha\equiv\alpha_{\phi\phi} h \etat^{-1}$, and $C_\Omega=s\Omega
h^2 \etat^{-1}$ (with $s\equiv{\rm d\,ln}\Omega/{\rm
  d\,ln}\,r$). Curiously, the resulting dynamo number $D \equiv
C_\alpha C_\Omega$ is asymptotically independent of $|\mB|$. In
practise, however, we have $q_\alpha>q_\eta$, which implies that
$\alpha$ is quenched earlier. Estimating $\tan(p) \simeq
(C_\alpha/C_\Omega)^{1/2}\propto |\mB|^{-1}$, we see that a quenched
$\alpha\Omega$~dynamo is dominated by differential rotation, leading
to vanishing radial pitch angle. The loss of significant pitch angle
in the saturated state can be circumvented if the dynamo is saturated
at high $C_\alpha$ -- e.g. via the vertical wind
\citep*[see][]{2009IAUS..259..467E}.


\section{Results}
\label{sec:results}

In the following, we will present results from a comprehensive suite
of simulations. After we have attempted to eliminate as many as
possible free parameters from our model, there remain only two major
aspects that require testing: (i) the disc mass, i.e., the central
input parameter governing the SF rate, and (ii) the initial topology
of the magnetic field. Moreover, we aim to study non-axisymmetric
modes, and whether the inclusion of the Navier-Stokes (NS) equation
has an effect on the evolution of the Galaxy as a whole. In
particular, we are interested in whether the MRI or convective
instabilities can emerge on scales long enough not to be immediately
affected by turbulent diffusion.

A compilation of the main results can be found in \Tab{tab:models},
where we also list the input parameters of our setup. Models labelled
with `s' are what we refer to as ``static'', i.e. the density and
velocity field are kept fixed during the evolution of the model. This
is commonly assumed for the ``kinematic'' dynamo problem, albeit one
of course includes a back-reaction of the field to obtain saturation
of the dynamo. For some of the 3D models, we furthermore evolve the
full MHD equations including the density and velocity field; these
simulations are labelled `d' for ``dynamic''. For our fiducial
axisymmetric (`X') model ``X1s'', we vary the disc mass in steps of
0.5 times the fiducial mass of $1.14\times 10^{10}\Msun$ (see first
four rows of \Tab{tab:models}). With models ``X2s-halo'' and
``X3s-VF'', we study the influence of a halo dynamo and vertical-field
seeding, respectively. The fiducial non-axisymmetric (`N') models
``N1s'' and ``N1d'' probe the effects of various seed-field
geometries. With the exception of the two models ``N2d-noD'' (without
any $\alpha$~effect, but including turbulent diffusion) and
``N2d-MRI'' (without \emph{any} prescribed turbulence effects), all
simulations include mean-field (MF) effects as described in
\Sec{sec:meanfield} above. Generally, the MF dynamo remains dominant
for these models, but subtle differences arise due to the long-term
evolution of, e.g., the density distribution, which indirectly enters
the MF prescription.


\begin{table*}
  \caption{Simulation parameters and results. \label{tab:models}}
  \begin{tabular}{p{1.5cm}cccccclcccccp{2.3cm}}
    \hline\\[-8pt]
    model & dim & MF & NS & halo & $M_{\rm gas}$ 
    & seed && parity & $p_{\rm in}$ & $p_{\rm out}$ & $\tau_{\rm e}$ 
    & $|\mB_{\rm sat}|$ &comments \\[2pt]
    & & & & & $\!\![10^{10}{\rm M_\odot}]\!\!$ & & & 
    & $[\,\degr\,]$ & $[\,\degr\,]$ & $[{\rm Gyr}]$ &$[\mu{\rm G}]$ & 
    \\[2pt]\hline\\[-8pt]


    X1s-0.5 & 2D & \yes & \no  & \no  & 0.57 & WN  &
    & S0/A0 & -11.6 &  -4.7 & 0.374 & 1.44 & 
    see \Fig{fig:bfield} \\

    X1s     & 2D & \yes & \no  & \no  & 1.14 & WN  &
    & S0$^a$\hspace{-1ex} & -11.0 &  -4.6 & 0.503 & 3.75 & 
    see Figs.~\ref{fig:pitch},$\,$\ref{fig:bsat_X1s}   \\

    X1s-1.5 & 2D & \yes & \no  & \no  & 1.70 & WN  &
    & S0$^a$\hspace{-1ex} & -10.8 &  -4.6 & 0.547 & 6.34 & \\

    X1s-2.0 & 2D & \yes & \no  & \no  & 2.27 & WN  &
    & S0     & -10.7 &  -4.3 & 0.593 & 9.07 & \\[4pt]


    X2s-halo & 2D & \yes & \no  & \yes & 1.14 & WN  &
    & S0$^a$\hspace{-1ex} & -11.7 &  -4.8 & 0.358 & 4.05 & \\

    X3s-VF & 2D & \yes & \no  & \no  & 1.14 & VF &
    & A0\ra S0 & -11.0 & -4.5 & 0.539 & 3.75 & \\


    \hline\\[-8pt]
    \multirow{2}{*}{%
    N1s/d-HF $\ \ \left<\frac{}{}\right.$} & 3D & \yes & \no  & \no  & 1.14 & 
    \multirow{2}{*}{HF} &
    & S0 & -11.0 &  -3.2 & -- &  3.75 & \\

              & 3D & \yes & \yes & \no  & 1.14 &     &
    & S0 & -~9.9 &  -2.6 & -- &  2.52 & \\[4pt]

    \multirow{2}{*}{%
    N1s/d-VF $\ \ \left<\frac{}{}\right.$} & 3D & \yes & \no  & \no  & 1.14 & 
    \multirow{2}{*}{VF, $\bar{B}_\phi$} &
    & S0 & -11.0 &  -6.1 & 0.409 & 3.75 & \\
 
              & 3D & \yes & \yes & \no  & 1.14 &     &
    & S0 & -~9.9 &  -2.7 & 0.407 & 2.65 & 
    see \Fig{fig:polmap}\\[4pt]
    

    N2d-noD & 3D & \no$^b$\hspace{-1ex}  & \yes & \no  & 1.14 & HF+VF &
    & A0 &  -1.7 &  -1.6 & -- & 0.82$^c$\hspace{-1ex} & 
    see \Fig{fig:bsat_comparison}\\

    N2d-MRI & 3D & \no & \yes & \no  & 1.14 & HF+VF &
    & A0 &  -0.4 &  -0.5 & -- & 4.25 & 
    see \Fig{fig:bsat_comparison}\\

    N3d-VF  & 3D & \yes & \yes & \no  & 1.14 & VF &
    & S0/A0 & -10.2 &  -3.4 & -- & 3.46 & 
    see Figs.~\ref{fig:bsat_comparison},$\,$\ref{fig:bfield_N3d},%
    $\,$\ref{fig:parker}\\[2pt]
    \hline
  \end{tabular}\\[6pt]
  \parbox[t]{2\columnwidth}{%
    \footnotesize
    $     ^a$ sub-dominant A0 outside $R\simeq 10\kpc$, 
    $\quad^b$ includes $\etat$, and $\vist$, 
    $\quad^c$ obtained outside $R\simeq15\kpc$.\\[6pt]
    All 2D runs are axisymmetric; mean-field (MF) effects include the
    ones described in Sect.~\ref{sec:meanfield}; runs including 'NS'
    evolve the Navier-Stokes equation. The `halo' dynamo is shown as a
    dashed line in \Fig{fig:ae_prof}. The column labelled $M_{\rm
      gas}$ gives the normalisation for the disc mass. For seed fields
    we use white noise (WN) with $0.15\nG$ rms
    amplitude, net-vertical field (VF) of $0.1\nG$, or
    net-horizontal field (HF) of $0.01\muG$. Pitch angles are given
    for the inner disc (i.e., where the magnetic field strength peaks)
    and for the outer disc (average for $R>10\kpc$) separately. Growth
    rates are for the magnetic field $|\mB|$, during an interval for
    which exponential growth can be identified.}
\end{table*}


To be as unrestrictive as possible on the emerging dynamo mode, we
generally apply a white-noise (WN) initial field, resulting in
approximately equal amounts of energy in all permissive modes.
Moreover, for the 2D models, we alternatively apply a net-vertical
field (VF), which might be hypothesised as a plausible seed
topology. Finally, in the 3D case, it furthermore becomes possible to
test a configuration with initially horizontal field (HF); even though
this topology is generally found to be impractical due to the
winding-up effect of the differential rotation
\citep{2012A&AT...27..319M}.

\subsection{Vertical parity and growth rates}
\label{sec:parity}

\begin{figure}
  \includegraphics[width=\columnwidth]{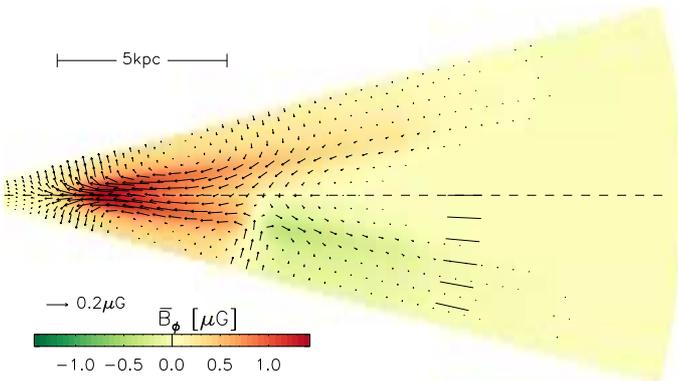}
  \caption{Saturated regular magnetic field for model X1s-0.5;
    colour-coded toroidal field, $\bar{B_\phi}$, overlaid vectors
    show the poloidal field. Short line segments indicate the
    latitudinal positions of slices in \Fig{fig:bsat_X1s}, below.}
  \label{fig:bfield}
\end{figure}

In agreement with many previous studies \citep[see
  e.g.][]{1996ARA&A..34..155B}, we find the axisymmetric (i.e., $m=0$,
hence ``0'') mode with symmetric vertical parity (``S'') as the
fastest growing dynamo mode (see column 8 in
\Tab{tab:models}). Notably, in the low-density part of the disc
(beyond $R\simeq10\kpc$), a weak anti-symmetric (``A'') mode
emerges. This is illustrated in \Fig{fig:bfield}, which shows the
saturated magnetic field for model X1s-0.5 (with lower disc mass),
where the A0 mode is seen most pronounced. Possible reasons for this
may be a combination of pumping and wind, as well as the disk
flaring. In most models, the A0 contribution remains sub-dominant, but
appears during a transitional phase in the case of a vertical-field
initial condition (e.g. model X3s-VF). Note that the mixed S0+A0 leads
to a radially confined, strong vertical field on only one side of the
Galactic disc. This is consistent with radio observations of
extragalactic Faraday rotation at high Galactic latitude by
\citet{2010ApJ...714.1170M}, who find a vertical field consistent with
$(0.00\pm 0.02)\muG$ towards the north, and $(0.31\pm0.03)\muG$
towards the south Galactic pole, respectively.

Exponential growth times of the mean-field dynamo are presented in
column 11 of \Tab{tab:models}. Generally, we find $\tau_{\rm e}$ on
the order of half a ${\rm Gyr}$, which is sufficient to explain the
present-day field strength of the Galaxy based on reasonable
assumptions on the initial seed field. \citet{2010Sci...328...73N}
estimated from Fermi observations a lower bound of of $3\times
10^{-16}\,{\rm G}$ for the intergalactic field on ${\rm Mpc}$
scales. A possible explanation of the generation process was recently
given by \citet{2012PhRvL.109z1101S}. The formation of the protogalaxy
leads to a further amplification up to $3\times 10^{-12}\,{\rm G}$
\citep{1995A&A...297..305L}, so roughly after $7 \Gyr$ the dynamo has
equipartition field strength of several microgauss. Prior to the epoch
of star formation, the MRI may grow unhindered by turbulence from SNe
and hence serve as a seed-field mechanism as suggested by
\citet{2004A&A...424..565K}.

For the fiducial model X1s, we find a trend to faster growth for
lighter disc models. This is presumably due to the reduced turbulent
diffusivity at lower SF activity. For lighter disc models we also
expect the wind to dominate over the vertical pumping, resulting in
weaker saturated fields and larger magnetic pitch angles. The fastest
growth of $\tau_{\rm e}=0.358\Gyr$ is found in model X2s-halo,
including a non-zero $\alpha$~effect at high Galactic latitude
(cf. dashed line in \Fig{fig:ae_prof}). Compared to the standard model
X1s, which is seeded from white noise, model X3s-VF with a
vertical-field initial condition shows a slightly slower growth rate
of $\tau_{\rm e}=0.539$, which can be identified with the A0
eigenmode. Faster growth can be obtained if one starts with the S0
mode directly -- cf. model N1s-VF, where a combined vertical and
azimuthal field is applied.

\subsection{ Radial structure}

The amount of quantitative information that can be confidently
extracted from radio observations of the Milky Way or nearby galaxies
is rather limited. In particular, reliable positional information is
restricted to considering the radial variation of azimuthally and
vertically averaged quantities. In the interest of direct quantitative
comparison with observations, we here want to discuss the magnetic
field strength and its inclination with respect to the toroidal
direction.

\subsubsection{The magnetic pitch angle}
\label{sec:pitch}

One major observable derived from radio polarisation maps is the
radial pitch angle, $p\equiv\tan^{-1}( \bar{B}_R/ \bar{B}_{\phi} )$ of
the mean magnetic field. Because values of $|p|\simgt 10\degr$ are
hard to explain in the presence of differential rotation alone, large
pitch angles are generally interpreted as the hallmark of an
$\alpha$~effect dynamo. Our models generally show moderately large
pitch angles close to $-10\degr$ (see column 9 of \Tab{tab:models}) in
the inner region of the Galactic disc, i.e. where the magnetic energy
is highest. The models N1d-HF/VF, including evolution of the disc,
show somewhat reduced pitch angles. This may be related to the
long-term viscous evolution of the radial gas density profile, which
reduces the SF rate entering our dynamo prescription, and hence the
strength of the MF dynamo near the Galactic centre. It is interesting
to note that we do not see a significant change of pitch angle during
the simulation -- implying a saturation at high $C_\alpha$, likely due
to the action of the vertical fountain flow
\citep[cf.][]{2009IAUS..259..467E}.

\begin{figure}
  \includegraphics[width=0.95\columnwidth]{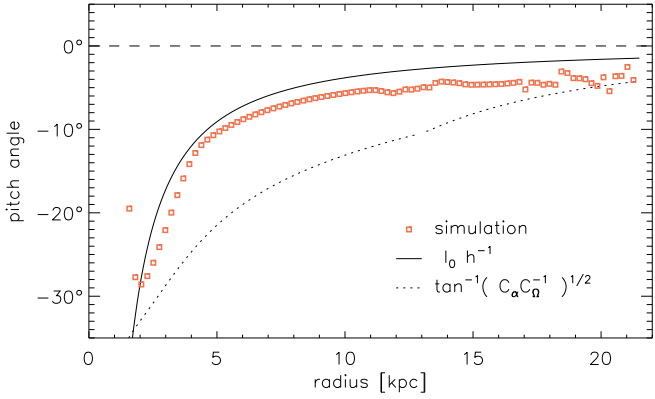}
  \caption{Radial pitch angle $p\equiv\tan^{-1}( \bar{B}_R/
    \bar{B}_{\phi} )$ for model X1s at the end of the simulation. The
    observed radial trend is well approximated by the crude estimate
    $p\simeq l_0\,h^{-1}$ \citep{2010ASP..conf..197F}, with
    $l_0=120\pc$.}
  \label{fig:pitch}
\end{figure}

\citet{2010ASP..conf..197F} points out the systematic variation of the
magnetic pitch angle with radius, which he ascribes to the flaring of
the disc. Similarly, \citet{2010ASPC..438..229R} have observed a
vanishingly small pitch angle for the outer Galaxy. In
Figure~\ref{fig:pitch}, we plot the radial profile of the pitch angle
(in the disc midplane) for the fiducial model X1s. The large angles of
the order of $-30\degr$ in the inner part should be ignored since the
mean field is significant below the equipartition value there (cf.
\Fig{fig:bsat_X1s} below), which should make them very hard to detect
in polarised synchrotron emission. Consistent with observations
\citep[see e.g. fig.~4 in][]{2010ASP..conf..197F}, the pitch angle in
our simulations decreases roughly like $R^{-1}$. Crudely estimating
$p\simeq l_0\,h^{-1}$, with a correlation length, $l_0$, of the
turbulence, this behaviour can conveniently be explained by the
flaring of the gas disc (cf. Fig~\ref{fig:disc_prof}). The good match
is somewhat deceiving as the supposedly more accurate estimate
\begin{equation}
  p \simeq \tan^{-1}\sqrt{C_\alpha\,C_\Omega^{-1}}
  \label{eq:pitch}
\end{equation}
in fact provides a much poorer description of the actual result.
Whereas \Fig{fig:pitch} shows the pitch angle in the Galactic
midplane, the peculiar shape of the dynamo A0 mode in the outer Galaxy
(seen in \Fig{fig:bfield} above) suggests to look at latitudinal
slices away from the midplane -- and, in fact, the pitch angle of the
A0 mode follows Eqn.~(\ref{eq:pitch}) more closely. Concluding this
section, we want to point out that the two models N2d-noD and N2d-MRI
without prescribed dynamo action produce vanishingly small pitch
angles. This is likely because the initial magnetic field is rather
small and accordingly the unstable modes lie close to the diffusive
border of the unstable region. In this case, the unstable mode is
characterised by a dominant toroidal field
\citep{2004A&A...424..565K}. The pitch angle is however larger for
model N3d-VF with the stronger initial field of $0.1\muG$.

\subsubsection{The saturated field strength}
\label{sec:bsat}

\begin{figure}
  \includegraphics[width=0.95\columnwidth]{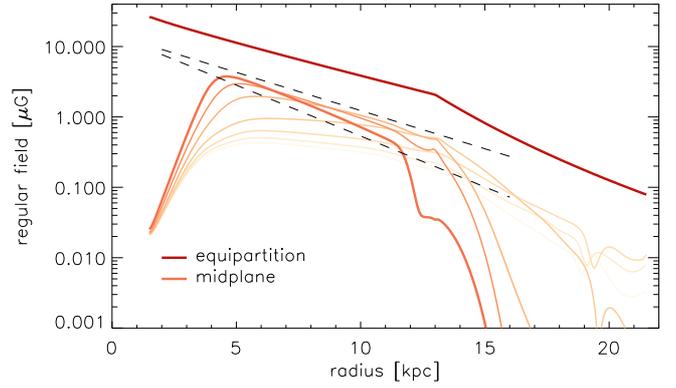}
  \caption{Saturated regular magnetic field for model X1s for various
    cuts of constant co-latitude. The midplane field peaks at about
    $4\muG$. Dashed lines show exponentials with scale lengths of 3
    and $4\kpc$, respectively.}
  \label{fig:bsat_X1s}
\end{figure}

Most of the nearby spiral galaxies are believed to be observed in the
saturated state of the dynamo \citep{1996ARA&A..34..155B}. With
increasingly detailed observations, it becomes possible to determine
the radial profile of the field strength \citep[see
  e.g.][]{2007A&A...470..539B}. Based on this, one can then estimate
the relative importance of magnetic forces on the overall rotational
balance within the Galactic disc
\citep[see][]{2013MNRAS.tmp.1539S}. Because we have started from a
quantitative disc model, we can hope to make meaningful predictions
about the radial distribution of magnetic field. Accordingly, in
\Fig{fig:bsat_X1s} we present radial profiles of the fiducial model
X1s in the saturated state. As one can see in \Fig{fig:bfield} above,
the dominant dynamo mode has a characteristic V-shape, which makes it
instructive to plot radial cuts at various angles $\theta$ away from
the midplane. The different curves are within $0-15\degr$ below the
midplane (in steps of $\delta_\theta=2.5\degr$) and cuts further away
from the midplane are shown in increasingly lighter colour. The
positions of the cuts are also indicated in \Fig{fig:bfield} by short
line segments. The profiles are steepest near the midplane, and
roughly follow exponential curves with scale lengths between $3-4\kpc$
as indicated by dashed lines in \Fig{fig:bsat_X1s}. This scale appears
to be partly inherited from the equipartition profile, which is itself
a consequence of the disc model and the scaling relations entering via
$\etat(r,\theta)$. Owing to the rather restrictive quenching factor of
$q_\alpha=10$, our dynamo solutions remain well below equipartition
strength (dark line), but peak values of a few microgauss are
nevertheless obtained (see column 12 in \Tab{tab:models}). The final
field strength shows substantial variation, illustrating the
dependence on the disc model. The highest absolute value (i.e.,
$9\muG$) of the mean field is found in the high disc mass case -- with
a trend to weaker fields for less massive gas discs. This trend also
explains the lower saturated field strength of $\simeq 2.5\muG$ in
model N1d-HF (and similarly N1d-VF) compared to $\simeq 3.8\muG$ in
N1s-HF (and N1s-VF), which does not include the evolution of the
disc's surface density.

\subsection{The role of dynamical instabilities}
\label{sec:mri}

One key aspect of the simulations presented in this paper is the
inclusion of the Navier-Stokes equations in the modelling, enabling us
to capture dynamical instabilities occurring on large length scales.
It has been suggested by \citet{1999ApJ...511..660S} that turbulence
created via MRI may play a role in the outer Galactic disc, i.e., in
the absence of significant star-formation activity and the associated
enhanced diffusion. In view of this, it is interesting to study the
hypothetical case of a Galactic disc without any star-formation
activity, for which one can perform simple MHD simulations without any
prescribed mean-field effects from unresolved scales \citep[see
  e.g.][]{2004A&A...423L..29D,0004-637X-764-1-81}. Such simulations
should however be interpreted with care since they neglect the
dominant source of energy input to the system, namely that from SNe.

\begin{figure}
  \includegraphics[width=\columnwidth]{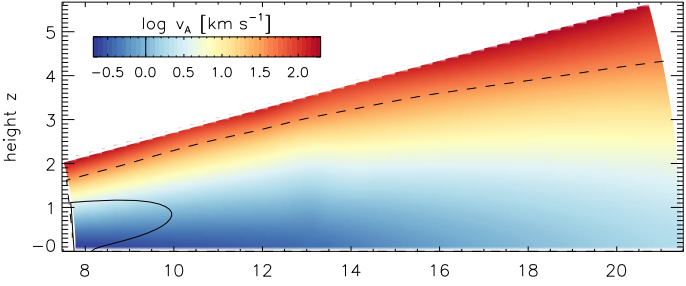}
  \caption{Vertical-field Alfv{\'e}n speed for the initial
    configuration of model N3d-VF (with $B_z=0.1\muG$), along with
    marginal stability lines according to the diffusive limit (solid)
    and strong-field limit (dashed line), represented by the left- and
    right-hand side of (\ref{eq:mri}), respectively.}
  \label{fig:mri}
\end{figure}

The occurrence of MRI -- subject to pre-existing turbulence from SNe
-- can easily be gauged from linear theory. According to the local
criterion derived in Appendix~A of \citet{2004A&A...424..565K}, who
study the linear stability of MRI in a global cylindrical disc of
semi-thickness $H$, instability is obtained within a range
\begin{equation}
  \sqrt{\frac{2-s}{s}}\ \frac{\etat}{H}\ \simlt \
  v_{\rm A}\ \simlt\ \sqrt{2 s}\ H\Omega\,,
  \label{eq:mri}
\end{equation}
where $v_{\rm A}\equiv |B_z|\,\rho^{-1/2}$ is with respect to the
vertical field. For a flat rotation curve with $s=1$, this yields
$\etat\,H^{-1} < v_{\rm A} < 1.4 H\Omega$, implying that already for
$C_\Omega > 1$, there exists a magnetic field unstable to MRI. This is
illustrated in \Fig{fig:mri}, where we show the marginal stability
lines for our model N3d-VF with a moderately strong initial field of
$0.1\muG$. For $R\simlt 8\kpc$, the region of possible MRI activity is
significantly restricted, whereas the outer disc clearly shows the
potential to develop the instability.

\begin{figure}
  \begin{center}
    \includegraphics[width=0.85\columnwidth]{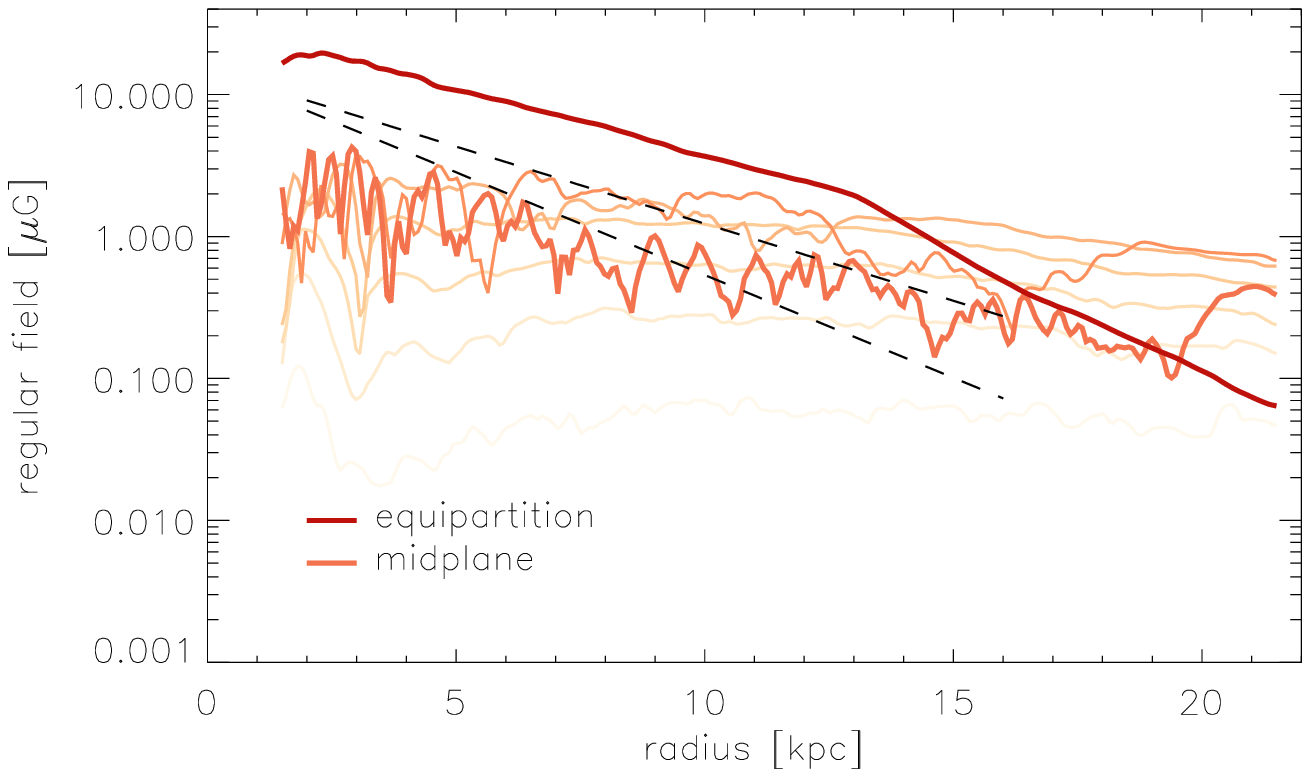}\\[2ex]
    \includegraphics[width=0.85\columnwidth]{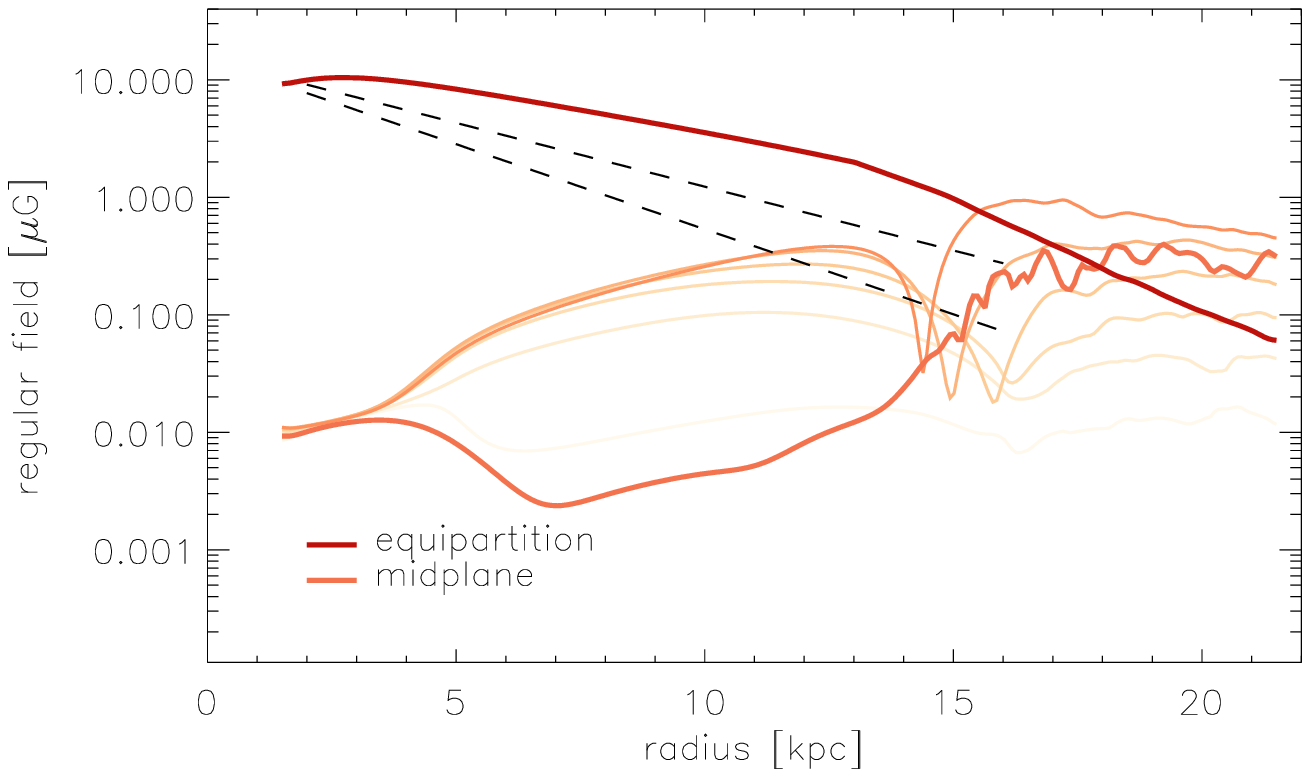}\\[2ex]
    \includegraphics[width=0.85\columnwidth]{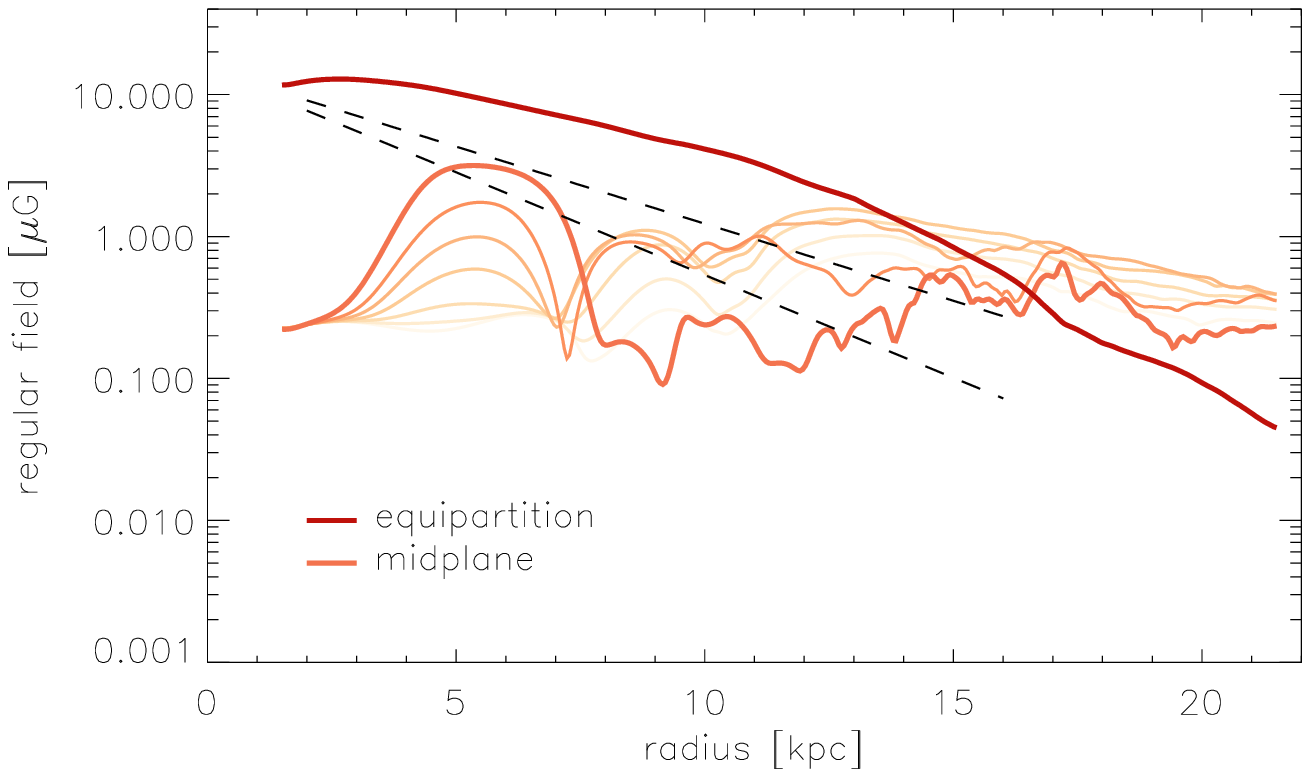}
  \end{center}
  \caption{Same as \Fig{fig:bsat_X1s} but for different dynamical
    models. \emph{Top panel:} Model N2d-MRI without prescribed MF
    effects; the radial scale length of the magnetic field is
    $\simeq10\kpc$. \emph{Middle panel:} Model N2d-noD without
    $\alpha$~effect but including turbulent diffusion (which
    suppresses the MRI for $R\simlt 15\kpc$). \emph{Bottom panel:}
    Model N3d with combined $\alpha$~effect and MRI, at $t=3.9\Gyr$,
    i.e., when the S0 mode dominates (cf. \Fig{fig:bfield_N3d}b).}
  \label{fig:bsat_comparison}
\end{figure}

In the three panels of \Fig{fig:bsat_comparison}, we try to illustrate
the interplay of prescribed small-scale effects with dynamical
instabilities. In the upper panel of that figure, we present model
N2d-MRI, where no dynamo-effects or turbulent diffusivity have been
prescribed, and where dynamo activity is due to magnetic instabilities
such as the MRI and convection. This case corresponds to the MHD
simulations of \citet{0004-637X-764-1-81}. In contrast to the static
dynamo simulation discussed earlier (see \Fig{fig:bsat_X1s}), it is
worth mentioning that in this case one obtains a much flatter radial
dependence of the mean magnetic field strength which is in better
agreement with observations \citep[see e.g.][which is however for the
  external galaxy NGC 6946]{2007A&A...470..539B}.

Because isolated MRI is a highly idealised scenario, we now turn to
the more realistic case where the MRI is affected by turbulence
resulting from SNe. In a general context, the effect of Ohmic
diffusion on growing MRI modes has been studied in the framework of
linear perturbation \citep{1996ApJ...457..798J}, and based on this,
\citet{2008AN....329..619G} have argued that the \emph{turbulent}
diffusion from SNe should be sufficient to damp the MRI at the solar
radius (also cf. \Fig{fig:mri} above). Whether simple picture of
``enhanced'' diffusion is viable needs to be scrutinised. Simulations
combining hydrodynamic forcing with the non-linear evolution of the
MRI \citep{2008ApJ...685..406W}, produce a rather varied
picture. However, for moderately strong small-scale forcing, the
authors indeed conclude that MRI may be suppressed by preexisting
turbulence. A similar conclusion is reached by
\citet{2010AN....331...34K}, who run shearing-box simulations with
finite Ohmic resistivity and, alternatively, with small-scale
forcing. The authors demonstrate that, assuming a typical $\etat(R)$,
MRI turbulence can be sustained inside the Galactic disc outside
$R\simeq 14\kpc$. This finding turns out to be in excellent agreement
with our model N2d-noD, where we apply turbulent diffusion (and
viscosity), but disable the mean-field dynamo. The resulting radial
field profiles are plotted in the middle panel of
\Fig{fig:bsat_comparison}.

As an aside, we note that, in MRI turbulence, the velocity dispersion
is typically found to be linked to the Alfv{\'e}n speed
\citep{2004A&A...423L..29D}. As can be seen in
\Fig{fig:bsat_comparison}, outside $\simeq15\kpc$ the MRI indeed leads
to super-equipartition field strengths with respect to the turbulent
kinetic energy input from SNe.\footnote{Note however that the
  corresponding line does not contain the turbulent kinetic energy
  created via the MRI itself.} Furthermore, in agreement with the
profiles obtained for NGC 6946 \citep{2007A&A...470..539B}, the radial
scale length of the regular field exceeds the scale length of the
turbulence -- this is unlikely for the dynamo-only case
(cf. \Fig{fig:bsat_X1s}).

Regarding the general field morphology (also see \Sec{sec:dynamical}
below), we find that anti-symmetric parity of the emerging mean field
prevails. This is despite linear theory predicts quadrupolar-like,
i.e. S0, parity \citep{2004A&A...424..565K} for MRI (albeit for a
non-flaring disc of constant thickness). Because of the dominance of
the A0 mode, the peak value of approximately $1\muG$ is reached away
from the disc midplane. Returning to the importance of pre-existing
turbulence, we want to emphasise that for model N2d-noD, turbulent
diffusion dominates in the inner disc (i.e., for $R\simlt 15\kpc$);
there the mean magnetic field remains at the initial seed-field
level. Away from the disc midplane, where the prescribed turbulent
diffusion is stronger, field is diffused-in from the MRI-active outer
region.

\begin{figure}
  \includegraphics[width=\columnwidth]{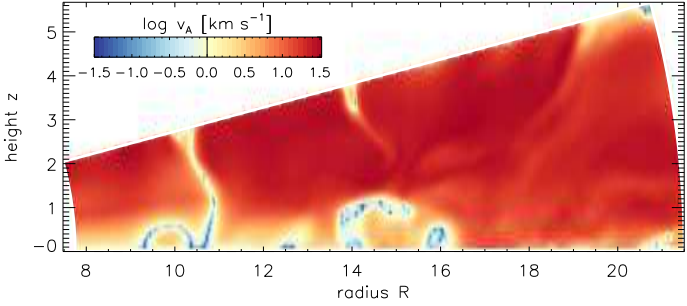}
  \caption{Alfv{\'e}n velocity based on the \emph{azimuthal} field,
    $\bar{B_\phi}$, for model N3d-VF at time $t=3.9\Gyr$. Velocities
    are surprisingly uniform reaching moderate values of several tens
    of $\rm km\,s^{-1}$.}
  \label{fig:alfven_N3d}
\end{figure}

Both of the two previously described scenarios only tell part of the
story. The models without any explicit mean-field effects (upper panel
of \Fig{fig:bsat_comparison}), and without $\alpha$~effect (middle
panel), instead should be contrasted with the case of a
\emph{combined} evolution of mean-field effects \emph{and} dynamical
instabilities on large scales -- this is shown in the lower panel of
the same figure. In the inner disc ($R\simlt 8\kpc$), the S0 dynamo
mode prevails, with strong fields near the midplane. In the range
$8\kpc \simlt R \simlt 10\kpc$, the A0 dynamo mode results in weak
fields near $z=0$, whereas for $R\simgt 14\kpc$ the magnetic field is
mostly due to MRI activity. Unlike in the case without prescribed
dynamo effects, at late times, i.e., after $t=3.9\Gyr$, the MRI now
reaches in to about $R\simgt 9\kpc$ -- probably assisted by
field-amplification via the $\alpha$~effect dynamo. Observationally,
the resulting radial scale length would likely appear considerably
larger than for a pure dynamo field (cf. \Fig{fig:bsat_X1s} above).

Returning to the question concerning effects of the regular magnetic
field on the Galactic rotation curve
\citep[see][]{2013MNRAS.tmp.1539S}, we point out that the relative
importance of magnetic fields in the overall force balance can be
roughly estimated by considering the strength of the field with
respect to the gas density. Accordingly, in \Fig{fig:alfven_N3d}, we
plot the azimuthal-field Alfv{\'e}n velocity for the third case
discussed above. Most notably, $v_{\rm A}$ is very uniform throughout
the outer Galactic disc. With values of several tens of $\rm
km\,s^{-1}$, the effect of the Lorentz force on the rotation curve is
likely to remain low and within current observational
uncertainties. Near the magnetic reversal at $R\simeq 10\kpc$, a minor
deviation of $\simeq +10\kms$ is seen in the rotational velocity at
intermediate Galactic latitude. Such a potential correlation between a
field reversal and a peak in the rotation curve may be a promising
future target for combined optical and radio-polarimetric
observations.

\subsection{Morphology of fully dynamical discs}
\label{sec:dynamical}

\begin{figure}
  (a)\\
  \includegraphics[width=\columnwidth]{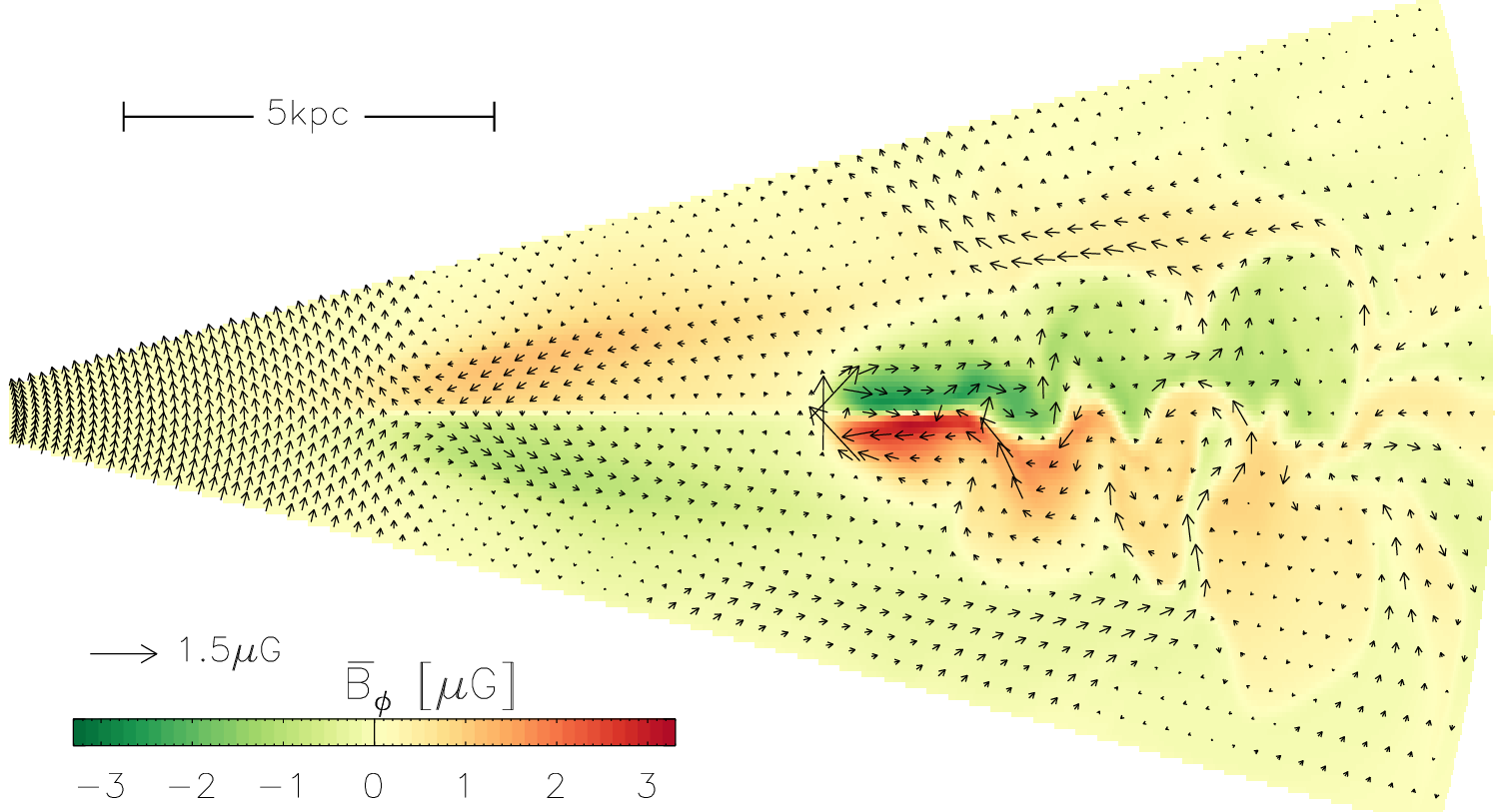}\\[2ex]
  (b)\\
  \includegraphics[width=\columnwidth]{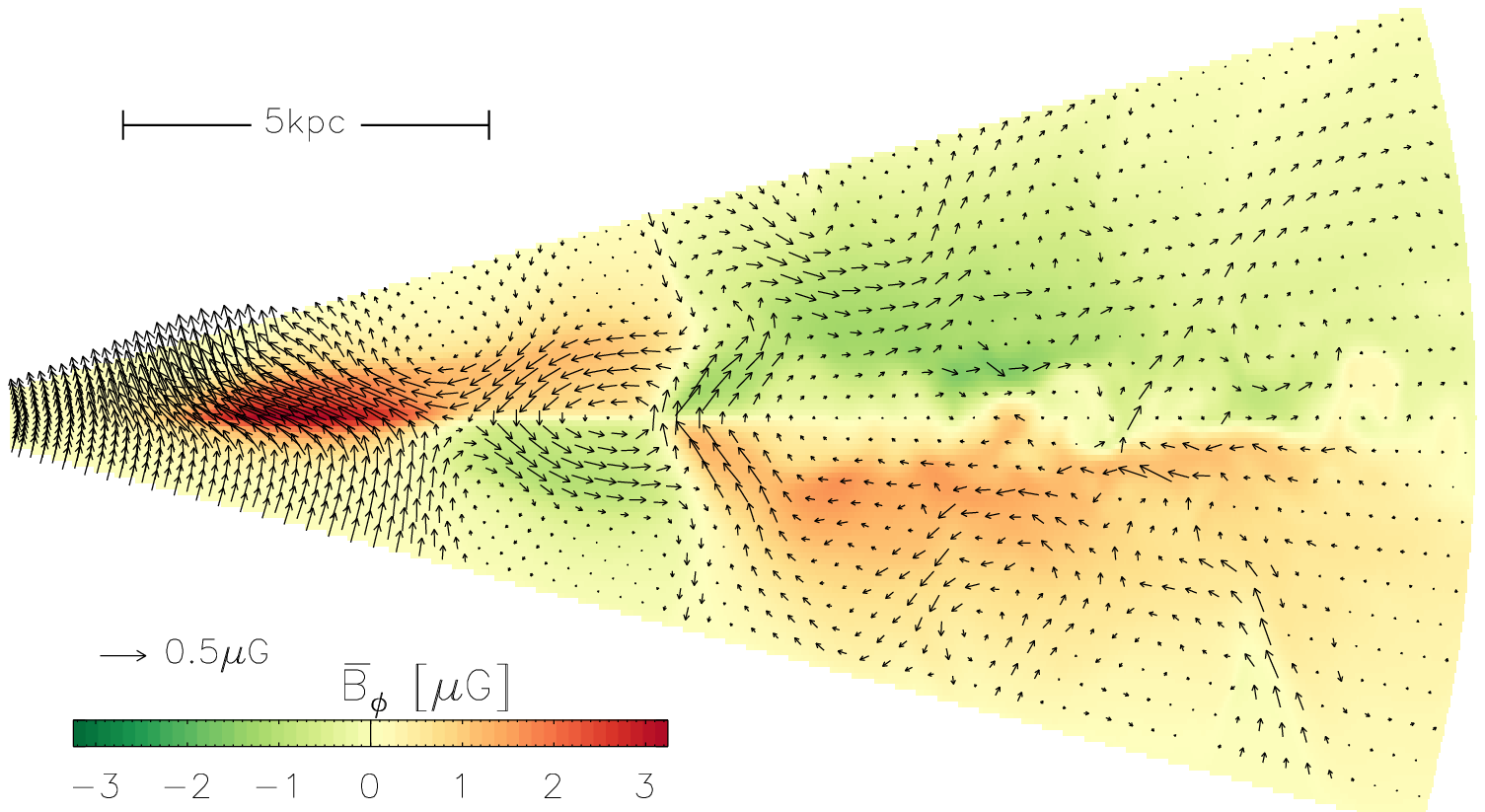}\\
  \caption{Poloidal cuts through model N3d-VF, with colour-coded
    $\bar{B}_\phi$ and vectors indicating the in-plane field. Panel
    (a), at $t=2.7\Gyr$, shows the initial A0 mode and strong fields
    created by the combined action of MRI and convection outside
    $R\simeq 10\kpc$. In panel (b), at $t=3.9\Gyr$, the S0 appears, and MRI is
    now somewhat weaker.}
  \label{fig:bfield_N3d}
\end{figure}

The emergence of the MRI in our simulations is limited by various
factors. Weak seed fields, for example, imply that the plasma
parameter $\beta_{\rm P}\equiv 2\bar{p}/\bar{B}^2$ falls into a range
where MRI only appears at high wavenumbers and only far up in the disc
where the gas pressure is low. To circumvent limitations due to
insufficient numerical resolution, we ran an additional scenario
``N3d-VF'', which is identical to N1d-VF, but has a higher initial
net-vertical magnetic field of $B_z=0.1\muG$, and adopts a higher
resolution of $384\times 72 \times 96$ grid cells. With this model, we
aim to study the combined effects of the mean-field dynamo and
secondary instabilities, as already illustrated in the previous
section.

A more detailed evolution of this model is presented in
\Fig{fig:bfield_N3d}, where we show vertical cuts through the domain.
Note that the $\bar{B}_\phi$ field created by the MRI (visible in the
outer disc) has the opposite polarity than the A0 dynamo mode (seen
between $5\kpc \simlt R \simlt 10\kpc $). This leads to a distinct
radius where the field is zero (which is also clearly visible in the
derived polarisation map -- see \Fig{fig:polmap} below). While this
would provide a natural explanation for the observed field reversal in
our own Galaxy, it is currently not clear whether the antagonism
between the MRI mode and the dynamo mode is random or systematic in
our model. A careful study of what determines the prevalent mode
structure in the presence of combined MRI and mean-field effects is
certainly called for.

Owing to the lower $\beta_{\rm P}$, the MRI develops closer to the
disc midplane. There the density is strongly stratified, which in turn
seems to lead to a Parker-type convective instability
\citep{1961PhFl....4..391N,1994A&A...287..297F,1995A&A...301..293F}
visible in the form of field arcs. Note that the vertical undulation
of the antisymmetric $\bar{B}_\phi$ creates apparent radial reversals
of the field direction in the disc midplane. It would be interesting
to check whether such a field distribution is consistent with all-sky
data of Faraday rotation and polarised emission \citep[see
  e.g.][]{2010MNRAS.401.1013J}. Clearly, this type of reversal would
be very difficult to observe in external galaxies -- which may
conveniently explain why the Milky Way appears unusual in this regard.

\begin{figure}
  \includegraphics[width=\columnwidth]{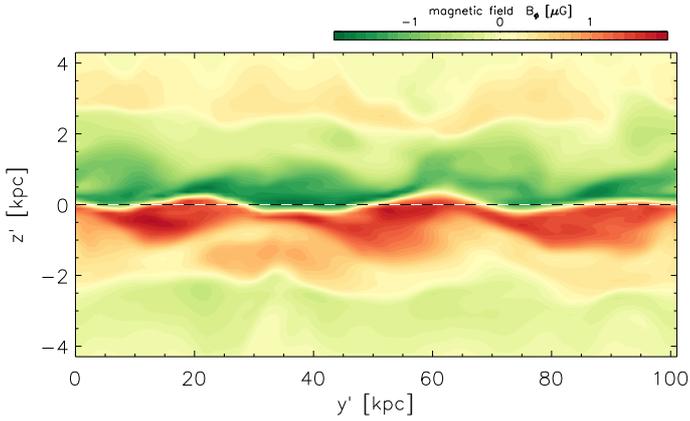}
  \caption{Projected $\theta$-$\phi$-slice of the azimuthal magnetic
    field at $r=16\kpc$. The dominant wavelength along the field lines
    is on the order of $30\kpc$.}
  \label{fig:parker}
\end{figure}

Our claim that a Parker-like buoyant instability is operating in the
regions of strong field creation has so far largely been guided by
visual appearance. In the following we hence try to assess, in a
semi-quantitative manner, the requirements for buoyancy instabilities
to occur. It is well established that the convective instability
\citep{1961PhFl....4..391N} works to interchange neighbouring segments
of field lines. Because magnetic tension forces oppose line bending,
the interchange will preferably occur on long segments. With dominant
$\bar{B}_\phi$ fields, one would thus expect perturbations that have
higher wave numbers in the radial direction compared to the azimuthal
direction. This is illustrated for model N3d-VF in \Fig{fig:parker},
where we show a $\theta$-$\phi$ slice of the $\bar{B}_\phi$ field
component (for the same point in time as for panel (a) in
\Fig{fig:bfield_N3d}). Note that \Fig{fig:parker} does not preserve
aspect ratio and the azimuthal coordinate $y'$ is strongly compressed
compared to the poloidal slice in the previous figure. The dominant
azimuthal mode appears to be $m=3$, which corresponds to a wavelength
of $\lambda_y\simeq 30\kpc$. In agreement with local simulations by
\citet{2008A&A...490..501J}, this is about a factor of 10--20 larger
than the radial wavelength $\lambda_x\simeq 2\kpc$. By order of
magnitude \citep{1961PhFl....4..391N}, one would expect $2\pi
L/\lambda_y \sim 1$, where $L$ is the scale height of the total
pressure $\bar{p}^{\star}\equiv \bar{p}+\bar{B}^2/2$. In agreement
with this estimate, we infer a value of $L\simeq5\kpc$. Based on our
simulation data, we moreover evaluate the basic stability criterion
for convective stability (assuming $\gamma=1$), the fastest growing
wave number ($\lambda_y\sim 10\kpc$), and growth rates ($\tau_{\rm e}
\sim 100\Myr$). While a quantitative comparison is hampered by the
fact that we cannot isolate the unperturbed background state, the
overall numbers support the notion that convective perturbations in
the disc are indeed created by magnetic buoyancy. We point out that
the inclusion of a non-isothermal equation of state may have a
stabilising effect on the system, which means that our current models
may overestimate related effects. On the other hand, this type of
instability has been argued to be enhanced by the presence of a
cosmic-ray component \citep{1992ApJ...401..137P,2004ApJ...605L..33H}.
As a concluding remark, we note that observational support for such
buoyant arcs remains largely unavailable -- although claims have been
made for magnetic loops in the inner Galactic disc
\citep{2006Sci...314..106F}.

\begin{figure}
  \center\includegraphics[width=0.9\columnwidth]{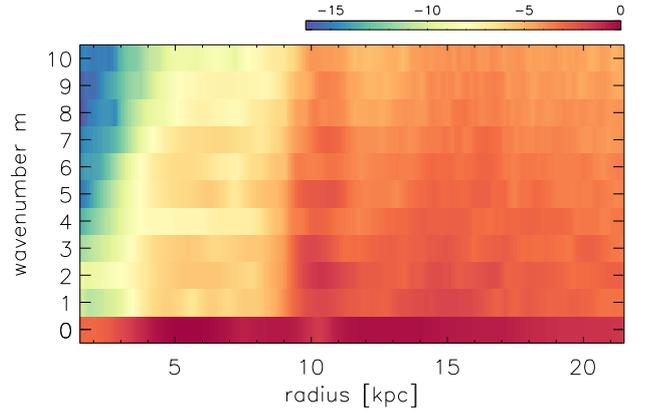}
  \caption{Logarithmic power spectrum of $\bar{B}_\phi$ as function of
    radius for model N3d-VF at time $t=3.9\Gyr$. The dynamo field is
    purely axisymmetric, whereas outside $R\simeq 9\kpc$ MRI
    turbulence is able to produce significant non-axisymmetric
    features.}
  \label{fig:mpower}
\end{figure}

To illustrate the very distinctive character of the dynamo-generated
field on one hand, and the field resulting from dynamical magnetic
instabilities on the other hand, in \Fig{fig:mpower}, we show the
vertically-averaged azimuthal power spectrum of the magnetic
field. The spectrum is taken from model N3d-VF at time $t=3.9\Gyr$,
and has been normalised to a maximum of one. The logarithmic colour
coding spans the available dynamic range of the double-precision
floating point numbers used in the simulation, ranging from
essentially zero (blue) to the maximum (dark red). The dynamo field is
characterised by a perfectly axisymmetric $m=0$ mode. Outside $R\simeq
9\kpc$, we find significant non-axisymmetric modes, but without any
particular mode number dominating. In the transition region between
the $\alpha\Omega$~dynamo and the MRI-dominated region, we see power
at odd overtones as well as a peak at $m=2$.


\begin{figure}
  \includegraphics[width=0.95\columnwidth]{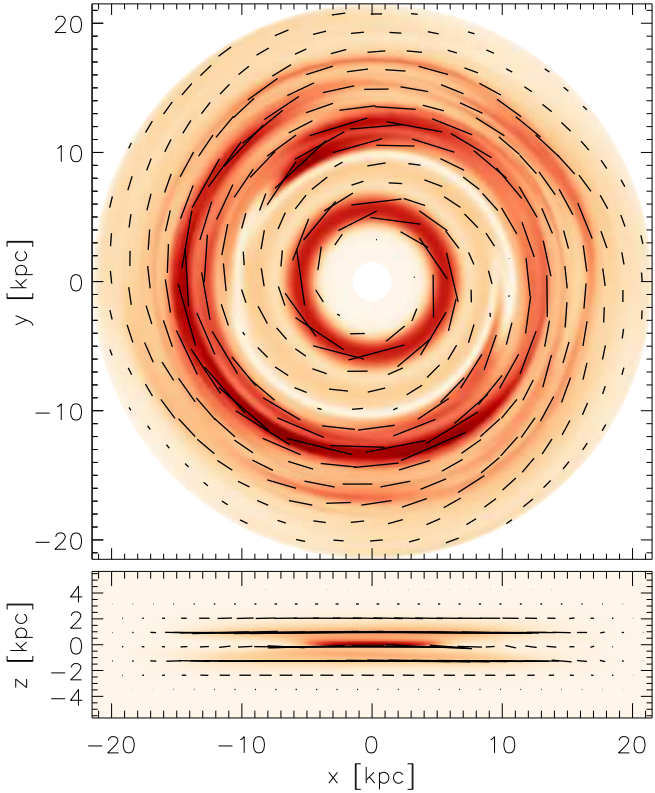}
  \caption{Synthetic polarisation maps for model N3d-VF at
    $t=3.9\Gyr$, i.e., corresponding to the lower panel in
    \Fig{fig:bfield_N3d}. Colour coding shows the Stokes-I parameter,
    and compass needles indicate the direction of polarisation
    (rotated by $90\degr$).}
  \label{fig:polmap}
\end{figure}

This feature is also seen in \Fig{fig:polmap}, where we show a very
basic polarisation map created from the saturated state of model
N3d-VF. This synthetic radio map integrates polarised synchrotron
emission along the line of sight, ignoring effects of Faraday-rotation
and -depolarisation, and assuming a flaring disc (of scale height
$h_{\rm rel}=1.5\kpc$ at $R=10\kpc$) for the density $n_{\rm rel}$ of
relativistic electrons. In the edge-on projection, we infer a radial
scale length of the total intensity of about $5\kpc$ if we assume a
radially constant $n_{\rm rel}$. If we assume a correlation $n_{\rm
  rel} \propto \bar{B}^2$, this is naturally reduced to
$2.5\kpc$. Needless to say that both results are consistent with the
radial scale length of $10\kpc$ seen for the magnetic field in
\Fig{fig:bsat_comparison}. We remark that these estimates refer to the
strong fields created in the outer disc by the MRI in combination with
buoyancy and the mean-field dynamo. We conclude that a potential field
reversal at a mode interface will make it difficult to infer a
meaningful global scale length from observations. This is in any case
difficult for the Milky Way itself, but slightly larger values have,
e.g., been obtained for NGC 6946 \citep{2007A&A...470..539B}, where a
scale length of $14\kpc$ has been found for the total magnetic field.

Ultimately, we aim to compare polarisation maps such as in
\Fig{fig:polmap} to observation-based models for the Galactic magnetic
field, e.g., the face-on view (top panel) of the heuristic model
derived by \citet{2012ApJ...757...14J}; their figure~9. We see that
our dynamo field is less centrally-confined than the best-fit to the
all-sky polarisation maps, and obviously lacks the precise spiral
features. More significantly, in the edge-on view (see lower panel of
\Fig{fig:polmap}), we do not see any X-shaped field topology. Even
though our dynamo models do produce significant vertical field, these
are always dominated by stronger radial and azimuthal fields. In the
edge-on projection, the polarisation vectors are accordingly aligned
with the horizontal direction.


\section{Summary of results}
\label{sec:summary}

With mean-field coefficients calibrated from direct SN simulations
(see \Tab{tab:coeff}), and with quenching functions determined
quantitatively \citep{2013MNRAS.429..967G}, we are left with
essentially no free input parameters other than the initial geometry
of the magnetic seed field. This of course precludes the possibility
to derive a bifurcation diagram for dynamo modes or critical dynamo
numbers \citep{1992A&A...259..453B}. On the other hand, it is
satisfying that, without tuning of any parameters, the outcome of our
simulations is indeed in decent agreement with observational
constraints.

\begin{itemize}

\item In all our models, we find a dominant S0 mode for the dynamo,
  but with a sub-dominant A0 mode situated in the outer disc. In the
  case of a low disc mass, the A0 mode is most pronounced. We moreover
  find anti-symmetric parity (notably of the opposite sign) for the
  dominant MRI mode.\medskip

\item The mixed S0+A0 dynamo mode leads to a localised region of
  strong vertical field, which is enforced by the requirement of a
  zero divergence. Because of the mixed parity, the vertical field
  only appears on one side of the disc, as is consistent with recent
  observations by \citet{2010ApJ...714.1170M}.\medskip

\item Consistent with a topical compilation of observations by
  \citet{2010ASP..conf..197F}, the radial profile of the magnetic
  pitch angle emerging from our model decreases with inverse
  radius. This is very well approximated by $p\simeq l_0 h^{-1}$, with
  $h$ the local scale-height of the flaring disc.\medskip

\item Vertical undulations caused by magnetic instabilities, in
  connection with an anti-symmetric vertical parity, can create
  apparent radial reversals near the disc midplane. The resulting
  field topology should be tested against available data of rotation
  measures in the Galactic plane. A reversal is also seen at the
  interface between the dynamo and MRI modes.\medskip

\item The most pronounced effect of allowing magnetic instabilities is
  to significantly enhance the radial scale length of the magnetic
  energy in the outer disc. Such shallow profiles, where the scale
  length of the field exceeds the one of the turbulence, have been
  observed in NGC 6946 \citep{2007A&A...470..539B}.\medskip

\item There has been renewed interest in the radial dependence of the
  magnetic field strength and its influence on the rotation curve
  \citep{2013MNRAS.tmp.1539S}. In our dynamo models, we find
  exponential scale lengths of $\sim 3-4\kpc$, which is somewhat
  shorter than expected. A shallower radial profile is seen in the
  disc halo, as well as in cases where the MRI leads to strong fields
  in the outer disc. Even there, deviations from the initial rotation
  curve stay well below current observational uncertainties.

\end{itemize}


\section{Conclusions}
\label{sec:conclusions}

We have described a new, comprehensive modelling approach for global
mean-field simulations of the Galactic dynamo. To be able to make
quantitative predictions, we aim to constrain all relevant input
parameters in a rigorous way. Our model for the gaseous disc is
derived in a self-consistent way, based on the observed gravitational
potential of the Galaxy, and its measured \HI distribution. The
prescription of mean-field effects (stemming from spatial scales
unresolved in the global simulation) is parametrised from a
comprehensive set of resolved shearing-box simulations including the
treatment of the multi-phase ISM and energy input from SNe.

In a further step towards improving the generality of our mean-field
models, we moreover solve the complete MHD equations rather than
keeping the flow field fixed. This is necessary to capture magnetic
instabilities like the MRI and convective instabilities. These arise
on scales large compared to the outer scale of the SN-driven
turbulence, where turbulent diffusion is less and less efficient. A
further prospect of solving the MHD equations is the future extension
of the model towards a self-consistent galaxy evolution model. Such a
model could potentially include spiral arm features (via self-gravity
of a stellar N-body population or the gas disc itself), effects from
accretion of material onto the galaxy, from galaxy encounters, or
ram-pressure stripping. As for all of these, the inclusion of
mean-field effects appears mandatory, as has recently been
demonstrated for the latter case by \citet{2012A&A...544A...5M}.

We stress that our model should only be regarded as a very first step
towards a fully comprehensive approach. Clearly there remain
discrepancies with respect to observations. For example, the only way
to produce X-shaped polarisation vectors in edge-on polarisation maps
is to already start with a strong vertical field (and to prescribe a
``differential'' wind in the radial direction). This is because in the
edge-on projection the horizontal disc field always dominates. We
hence conjecture that the X-shaped fields seen in edge-on galaxies are
possibly not the result of a disc dynamo. In contrast, dynamo
simulations that show X-like field configurations typically include
the polar regions of the spherical domain, something that is currently
missing in our own description. Examples include the work by
\citet{1993A&A...271...36B}, where also a much stronger wind is
assumed. More recently, \citet{2010A&A...512A..61M} have obtained
X-shaped topologies in simulations including an $\alpha$~effect in the
halo itself \citep[also cf.][]{2008A&A...487..197M}. Such a spherical
halo dynamo had originally been proposed by
\citet{1990Natur.347...51S}. In the absence of such an effect, a
strong outflow \citep[such as, e.g., seen in
  NGC253,][]{2009A&A...506.1123H,2011A&A...535A..79H} is probably
required to explain such field geometries. One way to explore this in
more detail will be global MHD simulations that incorporate both
mean-field effects from small scale turbulence and the effect of CRs
and SNe in form of a nuclear star-burst \citep{2013MNRAS.430.3235M}
and super-bubbles. Such simulations however must account for the
multi-phase nature of the ISM (via an appropriate cooling function),
and are highly demanding in terms of CPU resources.


\vspace{-1ex}
\begin{acknowledgements}
We thank the referee, David Moss, for useful comments that led to an
improvement of the paper.
This project is part of DFG research unit 1254. Computations were
performed on resources provided by the Swedish National Infrastructure
for Computing (SNIC) at the High Performance Computing Center North
(HPC2N) in Ume{\aa}.
\end{acknowledgements}
\vspace{-4ex}


\appendix


\end{document}